\documentclass[%
 reprint,
superscriptaddress,
 amsmath,amssymb,
aps,
prl
]{revtex4-2}

\DeclareMathOperator{\Tr}{Tr}% Added by user

\usepackage{graphicx}% Include figure files
\usepackage{dcolumn}% Align table columns on decimal point
\usepackage{bm}% bold math
\usepackage{hyperref}% add hypertext capabilities
\hypersetup{
    colorlinks=true,
    linkcolor=blue,
    citecolor=blue,     
    urlcolor=blue,
}% Added by user
\allowdisplaybreaks% Added by user

\begin{document}

\preprint{APS/123-QED}

\title{Hall Coefficient and Resistivity in the Doped Bilayer Hubbard Model}% Force line breaks with \\

\author{Yin Shi}
 \email{yxs187@psu.edu}
\affiliation{%
 Department of Materials Science and Engineering, The Pennsylvania State University, University Park, Pennsylvania 16802
}%
\author{Jonathan Schirmer}
\email{jzs429@psu.edu}
\affiliation{%
 Department of Physics, The Pennsylvania State University, University Park, Pennsylvania 16802
}%
\author{Long-Qing Chen}%
% \email{lqc3@psu.edu}
\affiliation{%
 Department of Materials Science and Engineering, The Pennsylvania State University, University Park, Pennsylvania 16802
}%


%\affiliation{
%\affiliation{
%\affiliation{%


\date{\today}% It is always \today, today,

\begin{abstract}
Finding and understanding non-Fermi liquid transport behaviors are at the core of condensed matter physics.
Most of the existing studies were devoted to the monolayer Hubbard model, which is the simplest model that captures essential features of high-temperature superconductivity.
Here we discover a new type of non-Fermi liquid behavior emergent in the hole-doped bilayer Hubbard model, using dynamical mean-field theory with a full consideration of the short-range interlayer electron correlation.
We find that at low temperatures, the Hall coefficient has a strong nonmonotonic dependence on temperature, leading to a double or quadruple reversal of its sign depending on the doping level.
At the same time, the resistivity exhibits two plateaus rather than linearity in its temperature dependence.
We show that these intriguing transport behaviors stem from the formation of coherent interlayer singlets, which scatter off gapped collective modes arising from short-range interlayer antiferromagnetic fluctuations.
%\begin{description}
%\end{description}
\end{abstract}

\maketitle

%\tableofcontents

\section{Introduction}
Studying the magnetotransport properties of electron systems is a valuable way to learn about their electronic structure.
For example, in high-temperature cuprate superconductors, the direct-current (DC) Hall resistivity has a strong temperature ($T$) dependence and changes its sign in the heavily overdoped regime~\cite{hwang94scaling,tsukada06negative}.
Meanwhile, the DC longitudinal resistivity in the normal state has linear temperature dependence and exceeds the Mott-Ioffe-Regel criterion~\cite{gurvitch87resistivity,ando04electronic,gunnarsson03colloquium}, known as strange metallicity.
In an atomically thin cuprate van der Waals heterostructure during cooling, the Hall resistivity decreases and changes from positive to negative and then reverses sign again before vanishing at low temperatures. This was explained by the vortex dynamics-based description of the Hall effect in high-temperature superconductors~\cite{zhao19sign}.
These behaviors are incompatible with the Fermi liquid theory of weakly interacting electrons and manifest the intricate nature of strongly correlated electron systems.

In efforts to understand the non-Fermi liquid behaviors, various authors have calculated the magnetotransport properties of the hole-doped Hubbard model using the quantum Monte Carlo method for small square lattices~\cite{assaad95hall}, the dynamical mean-field theory (DMFT) approximation for hypercubic~\cite{lange99magnetotransport} and square~\cite{markov19robustness,vucicevic21universal,kuchinskii22hall} lattices, and an expansion formula of the Hall coefficient for small square lattices~\cite{wang20dc}.
A double sign change of the $T$-dependent DC Hall coefficient similar to that in cuprate superconductors has been observed~\cite{lange99magnetotransport,wang20dc}.
Recent numerical calculations for the square-lattice Hubbard model also revealed the $T$-linear DC longitudinal resistivity exceeding the Mott-Ioffe-Regel limit~\cite{huang19strange} and a $T$-linear electron scattering rate at low temperatures~\cite{wu22non}.

These works motivate us to investigate further the magnetotransport properties of a more complicated lattice model, the Hubbard model on a bilayer square lattice, in which electrons can form disordered interlayer singlets with a spin gap~\cite{najera17resolving,najera18multiple,mou22bilayer}.
Accurately computing the conductivities of strongly correlated systems is notoriously difficult~\cite{auerbach18hall}, and is frequently hindered by small lattice sizes, infinite expansion summations, or the omission of vertex corrections.
We use the DMFT~\cite{georges96dynamical} to calculate the resistivities of the hole-doped bilayer Hubbard model.
The DMFT works for the thermodynamic limit, but in this theory the vertex corrections to the in-plane conductivities cancel out due to the neglect of in-plane nonlocal correlations~\cite{vucicevic21electrical}.
However, the short-range out-of-plane correlation is entirely accounted for in the Kubo bubble. This distinguishes our calculation from those for the monolayer Hubbard model.

We find that the Hall coefficient has a strong nonmonotonic $T$ dependence at low temperatures and can change its sign twice or four times with decreasing temperature, depending on the doping level.
Concomitantly, the longitudinal resistivity as a function of $T$ acquires two plateaus that smoothly cross over to each other.
These unfamiliar transport behaviors are shown to be associated with the formation of coherent interlayer singlets, which scatter off gapped collective modes arising from short-range interlayer antiferromagnetic fluctuations.

\begin{figure*}
\includegraphics{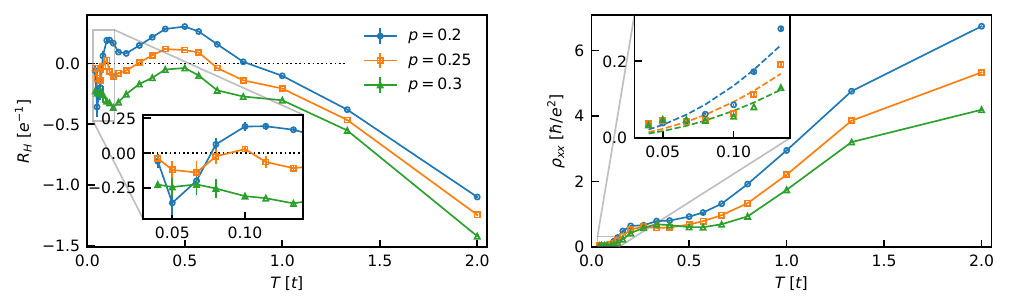}% Here is how to import EPS art
\caption{\label{fig:resis}Hall coefficient (left panel) and longitudinal resistivity (right panel) of the bilayer Hubbard model as functions of the temperature at various doping levels.
%The color shades in the left panel mark different regimes defined by the $T$ dependence of $R_H$ (see text).
The insets are close-up views of the lowest-temperature data.
The solid lines are the guide to the eye. 
The dashed lines in the inset of the right panel are quadratic fits $\rho_{xx}=\mathrm{const.}\times T^2$. 
The error bars represent Monte Carlo sampling errors and errors arising from DMFT iterations~\cite{gull07performance}, determined by four iterations starting from a converged solution.}
\end{figure*}

\section{Model and methods}
The bilayer square-lattice Hubbard model consists of two square lattices stacked site-to-site. 
We consider only the nearest-neighbor intralayer hopping $t$ and interlayer hopping $t_\perp$.
%The orbital effect of a uniform magnetic field $B_z$ perpendicular to the lattice plane is to give $t$ a Peierls phase, $\theta_{\imath\jmath}=\pi(y_\jmath-y_\imath)(x_\imath+x_\jmath)B_z a^2/\Phi_0$ in the Landau gauge~\cite{vucicevic21electrical}.
%Here $(x_\imath,y_\imath)$ is the in-plane site coordinates in units of the in-plane lattice constant $a$.
%$\Phi_0=h/e$ is the magnetic flux quantum.
The Hamiltonian is
\begin{eqnarray*}
H &=& -\sum_{\ell=1}^2 \sum_{\langle \imath,\jmath \rangle,\sigma} t c_{\ell \imath \sigma}^\dagger c_{\ell \jmath \sigma} - \sum_{\imath, \sigma} (t_\perp c_{1\imath \sigma}^\dagger c_{2\imath \sigma} + \text{H.c.})  \nonumber  \\
& &+ \sum_{\ell=1}^2 \sum_\imath U n_{\ell \imath \uparrow} n_{\ell \imath \downarrow}.
\end{eqnarray*}
Here $c_{\ell \imath \sigma}$ ($c_{\ell \imath \sigma}^{\dagger}$) annihilates (creates) an electron of spin $\sigma$ ($=\uparrow,\downarrow$) on the site $\imath$ in the layer $\ell$.
$U$ is the onsite Coulomb repulsion and $n_{\ell \imath \sigma}=c_{\ell \imath \sigma}^\dagger c_{\ell \imath \sigma}$ is the electron number operator.
%To facilitate calculations, $B_z$ is chosen to be in the form $B_z a^2/\Phi_0=p/q$ with $p$ and $q$ being coprime integers to obtain commensurate magnetic unit cells containing $q$ zero-field unit cells along the $x$ direction~\cite{markov19robustness,vucicevic21electrical}.
%We do not include the Zeeman splitting effect, just focusing on the orbital effect of magnetic fields.
The Kubo formulae for the longitudinal and Hall conductivities (sheet conductances) in a vanishing out-of-plane magnetic field $B_z\rightarrow 0$ can be directly extended from those for the monolayer Hubbard model~\cite{lange99magnetotransport,voruganti92conductivity,pruschke95anomalous},
\begin{eqnarray*}
\sigma_{xx} &=& \frac{e^2\pi}{\hbar N} \sum_{\mathbf{k},\sigma} \left( \frac{\partial \epsilon_\mathbf{k}}{\partial k_x} \right)^2 \int d\omega \Tr[ \hat{A}_{\mathbf{k}\sigma}(\omega)^2 ] \left[-\frac{df(\omega)}{d\omega}\right],  \\
\frac{\sigma_{xy}}{B_z} &=& \frac{2\pi^2e^3a^2}{3\hbar^2 N} \sum_{\mathbf{k},\sigma} \left( \frac{\partial \epsilon_\mathbf{k}}{\partial k_x} \right)^2 \frac{\partial^2 \epsilon_\mathbf{k}}{\partial k_y^2}  \\
& &\times \int d\omega \Tr[ \hat{A}_{\mathbf{k}\sigma}(\omega)^3 ] \left[-\frac{df(\omega)}{d\omega}\right].
\end{eqnarray*}
Here $e$ is the elementary charge magnitude, $\hbar$ is the reduced Planck constant, $a$ is the in-plane lattice constant, $N$ is the number of unit cells in the lattice, and $\mathbf{k}=(k_x, k_y)$ is the reciprocal vector in the first Brillouin zone.
$f(\omega) = (1+e^{\hbar \omega/T})^{-1}$ is the Fermi distribution function.
The energy of the bonding or antibonding band up to a constant shift is $\epsilon_\mathbf{k}=-2t(\cos k_x + \cos k_y)$.
$\hat{A}$ is the spectral function, which is a matrix in the layer index space.

We choose $t_\perp=1.2t$ and $U=10t$, which are relevant to the material VO$_2$~\cite{najera17resolving}, a prototypical strongly correlated oxide with the vanadium dimer as the basic unit~\cite{biermann05dynamical}.
In the DMFT, we consider an interlayer dimer embedded in a self-consistent noninteracting electron bath, thereby fully taking into account the short-range interlayer electron correlation.
We use the continuous-time auxiliary-field Monte Carlo method~\cite{gull08continuous,gull11continuous} to solve the corresponding quantum impurity problem, accurately measuring the self-energy at all Matsubara frequencies.
We then employ the recently developed maximum quantum entropy method~\cite{sim18maximum,bergeron16algorithms} to analytically continue the self-energy matrix to the real-frequency axis.

\section{Transport coefficients}
Figure~\ref{fig:resis} shows the calculated Hall coefficient $R_H=\sigma_{xy}/(B_z \sigma_{xx}^2)$  and longitudinal resistivity $\rho_{xx}=\sigma_{xx}^{-1}$ as functions of the temperature at various hole doping levels $p=1-\sum_\sigma\langle n_{\ell\imath\sigma} \rangle$.
For $T\gtrsim 0.1t$, the $T$ dependence of $R_H$ is similar for all doping levels shown, but $R_H$ shifts downward with increasing doping.
In this temperature range, as $T$ increases, $R_H$ decreases in $0.1t \lesssim T \lesssim 0.13t$, then increases in $0.13t \lesssim T \lesssim 0.5t$, and then decreases again for $T \gtrsim 0.5t$.
In $0.67t  \lesssim T \lesssim 1t$, $R_H(T)$ changes more slowly with increasing doping and becomes almost flat at $p=0.3$.
Depending on the doping level, $R_H$ can be totally below zero ($p=0.3$), or change sign once ($p=0.2$) or three times ($p=0.25$) in this range $T \gtrsim 0.1t$.

For $T \lesssim 0.1t$, the behaviors of $R_H(T)$ at different doping levels are radically different.
In this temperature range, the $T$ dependence of $R_H$ quickly weakens with increasing doping.
At $p=0.2$ and $p=0.25$, $R_H$ changes sign once due to its strong dependence on $T$.
But for a heavier doping $p=0.3$, $R_H(T)$ is almost a negative constant.
Therefore, the total number of times $R_H(T)$ changes its sign counts to zero at $p=0.3$, two at $p=0.2$, and as many as four at $p=0.25$, in contrast to the single or double sign reversal normally observed in high-temperature superconductors~\cite{tsukada06negative,li19superconductivity,zhao19sign} and the single-orbital Hubbard model~\cite{lange99magnetotransport,wang20dc}.

\begin{figure*}
\includegraphics{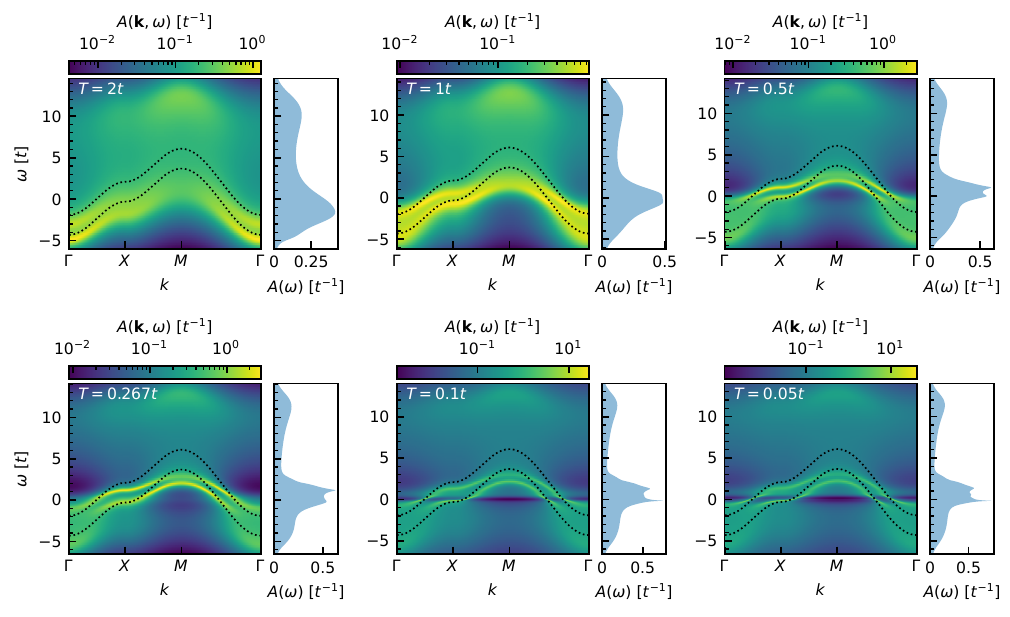}% Here is how to import EPS art
\caption{\label{fig:spectra}Single-particle excitation spectra $A(\mathbf{k},\omega)$ along a path connecting high-symmetry reciprocal points, and corresponding densities of states $A(\omega)$ at indicated temperatures for $p=0.25$. The black dotted lines are the noninteracting bands for the same doping. The Fermi level lies at $\omega=0$.}
\end{figure*}

The longitudinal resistivity $\rho_{xx}(T)$ also shows unfamiliar behavior (Fig.~\ref{fig:resis}, right panel).
There are two temperature ranges, $0.26t \lesssim T \lesssim 0.5t$ and $T \lesssim 0.1t$, where $\rho_{xx}(T)$ is almost constant, and these ranges become broader for heavier doping.
Especially at low temperatures, $\rho_{xx}(T)$ deviates significantly from the quadratic fits $\rho_{xx}=\mathrm{const.}\times T^2$ expected for a Fermi liquid (Fig.~\ref{fig:resis}, right panel, inset).
Nevertheless, the quadratic fit is improved for heavier doping, along with the weaker $T$ dependence of $R_H$ for heavier doping (Fig.~\ref{fig:resis}, left panel, inset), demonstrating that the system at low temperatures approaches the Fermi liquid phase as doping increases.
The constant value of $\rho_{xx}$ in $T\lesssim 0.1t$ does not change much as the doping level is varied, whereas at high temperatures, $\rho_{xx}$ exceeds the Mott-Ioffe-Regel limit ($\sim \sqrt{2\pi}\hbar/e^2 \approx 2.5 \hbar/e^2$~\cite{gunnarsson03colloquium}) and is lower for heavier doping consistent with more charge carriers.

\section{Mechanisms}
To understand the anomalous behaviors of the Hall coefficient and longitudinal resistivity, we plot in Fig.~\ref{fig:spectra} the single-particle excitation spectra $A(\mathbf{k},\omega)=\sum_\sigma \Tr \hat{A}_{\mathbf{k}\sigma}(\omega)$ and densities of states $A(\omega)=\sum_\mathbf{k}A(\mathbf{k},\omega)/N$ at various temperatures for $p=0.25$.
The noninteracting band structure is also superimposed (dotted lines).
In the noninteracting limit, at light doping, the bonding (lower-lying) band has a hole pocket at the $M$ point and the antibonding (higher-lying) band has a smaller electron pocket at the $\Gamma$ point, which is the case for $p=0.25$.
At heavy doping, both the bonding and antibonding bands have an electron pocket at the $\Gamma$ point.

At a high temperature $T=2t$, the spectrum is highly incoherent and continuous, showing only two broad Hubbard bands centered around $\omega=-1t$ and $\omega=10t$, respectively, with a pseudogap in between.
Charge excitation across this pseudogap produces a negative $R_H$~\cite{lange99magnetotransport}.
As $T$ is lowered to $1t$, the spectrum near the Fermi level becomes more coherent, and the peak of the density of states of the lower Hubbard band moves to a higher energy.
Fewer electrons are excited onto the upper Hubbard band, leading to a more holelike $R_H$, that is, a larger $R_H$.

\begin{figure*}
\includegraphics{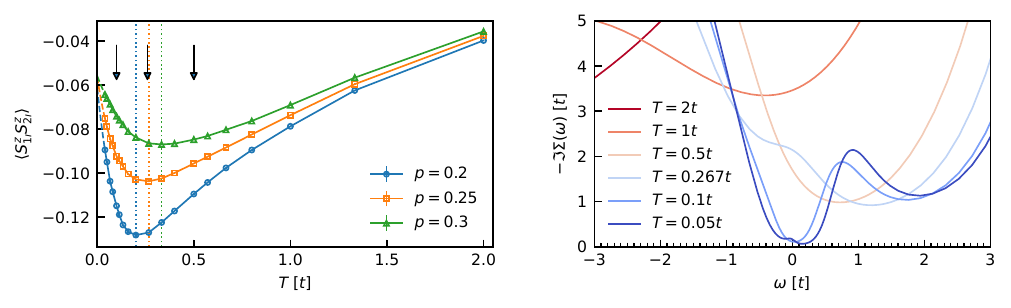}% Here is how to import EPS art
\caption{\label{fig:singlet}Left panel: interlayer singlet correlation as a function of the temperature at various doping levels. The vertical dotted lines mark the temperatures of the maximum correlation strength, $T_m$. The dashed lines are extrapolations to zero temperature by fitting the data for $T<T_m$ to cubic functions. The arrows indicate the approximate boundaries of the two resistivity plateaus. Right panel: imaginary part of the self-energy in the low-energy range at $p=0.25$ for the various temperatures in correspondence to Fig.~\ref{fig:spectra}.}
\end{figure*}

At a moderate temperature $T=0.5t$, two dispersive quasiparticle bands develop near and above the Fermi level, whereas the two incoherent Hubbard bands persist.
%The quasiparticle bands are formed by interlayer singlets on the lattice, as demonstrated in the infinite-coordination dimer Hubbard model at half-filling~\cite{najera17resolving}.
The quasiparticle bands are renormalized by the Kondo screening~\cite{najera17resolving} to be approximately twice as narrow as their noninteracting counterparts.
To see the interlayer Kondo screening, we show in the left panel of Fig.~\ref{fig:singlet} the nearest-neighbor interlayer spin correlation function $\langle S_{1\imath}^z S_{2\imath}^z \rangle$ as a function of temperature at various doping levels, where $S_{\ell\imath}^z=n_{\ell\imath\uparrow} - n_{\ell\imath\downarrow}$ is the spin density operator.
The neighboring interlayer spins do tend to be antiparallel screening each other.
The singlet correlation strength peaks at a nonzero temperature that increases with increasing doping.
For $T \lesssim 0.5t$, the singlet correlation strength is significant~\cite{wu22non} and considerably larger than that at $T=2t$, compatible with the observed crossover from the totally incoherent spectrum at $T=2t$ to the coherent bands at $T=0.5t$, which can thus be viewed as interlayer singlet bands~\cite{najera17resolving}.
Compared to the spectrum at $T=1t$, the hole pocket at the $M$ point is relatively well defined thanks to the coherent singlet bands, while the electron pocket at the $\Gamma$ point remains incoherent, leading to an overall positive $R_H$ (Fig.~\ref{fig:resis}, left panel).

%At $T=0.5t$, the \emph{incoherent} spectral weight around the Fermi level ($\sim 0.01t^{-1}$) is much reduced compared to that at $T=2t$ ($\sim 1t^{-1}$).
As $T$ is lowered from $0.5t$ to $0.267t$, the spectral function does not change much; especially the smearing of the quasiparticle bands at these two temperatures is similar, both being $\sim 0.5t$ wide and having a spectral weight of $\sim 3t^{-1}$.
This is consistent with the Kondo resonance, in which the width and height of the resonance peak are temperature-independent and are only determined by one energy scale, the Kondo temperature~\cite{mahan00many}.
%The Kondo resonance that takes place in $0.26t \lesssim T \lesssim 0.5t$ leads to the formation of interlayer singlets, so it should arise from the interlayer exchange interaction approximately proportional to $t_\perp^2$.
In this temperature range, $T$ is less than the smearing width of the quasiparticle bands, resulting in approximately $T$-constant resistivity (Fig.~\ref{fig:resis}, right panel) reminiscent of resistivity saturation in the Kondo problem.
There are two slight, yet visible changes in the spectral function as $T$ is lowered from $0.5t$ to $0.267t$.
One is that the electron pocket at the $\Gamma$ point becomes more well defined and deeper on the way to the complete formation of the coherent singlet bands.
The other is the appearance of a weak incoherent flat band with a smearing width of $\sim 1t$ at the Fermi level, which hosts electrons.
These two changes tend to reduce $R_H$ (Fig.~\ref{fig:resis}, left panel).

At a relatively low temperature $T=0.1t$, a small single-particle gap \emph{in the incoherent spectrum} ($\sim 0.4t$) opens at the Fermi level.
Concomitantly, the coherent singlet bands within the incoherent spectrum gap (spectral weight $\sim 30t^{-1}$) become much sharper and flatter than those at $T=0.267t$ (spectral weight $\sim 3t^{-1}$). 
This results in a very sharp peak in the density of states at the Fermi level, which is the manifestation of a strong Kondo resonance.
%This Kondo resonance should arise from the intralayer exchange interaction with bath electrons approximately proportional to $t^2 < t_\perp^2$, thus occurring at a lower temperature than the interlayer Kondo resonance.
The opening of the gap in the incoherent spectrum depopulates the upper Hubbard band, and the hole pocket of the well-defined singlet bands is larger than its electron pocket, resulting in an increase in $R_H(T)$ from $T\simeq 0.13t$ to $T\simeq 0.1t$ (Fig.~\ref{fig:resis}, left panel).

As $T$ decreases from $0.1t$ to $0.05t$, the quasiparticle bands again remain almost unchanged, with a $\sim 0.1t$ smearing width and $\sim 30t^{-1}$ spectral weight at the Fermi level.
Therefore, the resistivity in $T \lesssim 0.1t$ should also be approximately temperature-independent (Fig.~\ref{fig:resis}, right panel).
The two plateaus in $\rho_{xx}(T)$ are smoothly connected by a crossover spanning from $T\simeq 0.1t$ to $T\simeq 0.26t$.
At $T=0.05t$, a flat band (not the quasiparticle band) with weak intensity ($\sim 0.1t^{-1}$) appears at the Fermi level, which hosts electrons leading to a decrease in $R_H(T)$ from $T\simeq 0.1t$ to $T\simeq 0.05t$ (Fig.~\ref{fig:resis}, left panel).

The non-Fermi liquid behavior down to very low temperatures also has a manifestation in the interlayer spin correlation, which does not vanish at zero temperature (Fig.~\ref{fig:singlet}, left panel).
Instead, the singlet correlation extrapolated to zero temperature is $\sim -0.06$, which is very close to the nearest-neighbor spin correlation responsible for the non-Fermi liquid scattering in the overdoped monolayer square-lattice Hubbard model at low temperatures~\cite{wu22non}.
A dynamical cluster approximation study of the doped bilayer Hubbard model suggested that there exists non-Fermi liquid behavior even in the absence of finite scattering rate at vanishing temperature, attributed to short-range interlayer antiferromagnetic fluctuation~\cite{lee14competition}.
This observation is consistent with our results, nevertheless we showed that the resulting non-Fermi liquid behavior is not the $T$-linear resistivity but a resistivity plateau.

%The lower plateau at $T \lesssim 0.1t$ is much flatter than the higher plateau at $0.26t \lesssim T \lesssim 0.5t$, which results from the extra protection from inelastic scattering provided by the incoherent spectrum gap.
%Concomitant with the crossover in $\rho_{xx}(T)$, $R_H(T)$ reaches a minimum that connects the two maxima corresponding to holelike transport for $p=0.25$.
%These crossovers are signatures of the opening of the gap in the incoherent spectrum and its concomitant resonance of the coherent bands.

To show where the Kondo saturation comes from, we depict in the right panel of Fig.~\ref{fig:singlet} the imaginary part of the self-energy, $\Sigma(\omega)=\sum_\sigma \Tr \hat{\Sigma}_\sigma(\omega)$, where $\hat{\Sigma}_\sigma(\omega)$ is the self-energy matrix of spin $\sigma$ in the layer-index space.
At the lowest temperature shown, $T=0.05t$, $-\Im \Sigma(\omega)$ has two peaks in the low-energy range located at $\omega \simeq 0$ and $\omega \simeq 0.9t$, respectively, which are separated by a gap spanning $0.05t \lesssim \omega \lesssim 0.3t\equiv \omega_g$.
These two peaks are absent in the DMFT result of the single-orbital Hubbard model [following the Fermi liquid behavior at low temperatures, $-\Im\Sigma(\omega)\sim\omega^2$ at small $\omega$~\cite{pruschke95anomalous,georges96dynamical}] and represent two low-energy scattering modes arising from short-range interlayer antiferromagnetic fluctuations.
The zero-energy mode should be the simultaneous flip of spins in the inert interlayer singlet costing no energy, while the second mode should be the flip of a single spin in the singlet.
There is also a pseudogap at $\omega\simeq 2t$ that separates the second scattering mode and a broad peak at a high energy $\omega\simeq 7.8t$ (not shown) corresponding to scattering off doubly-occupied-site states.

The zero-energy scattering mode is responsible for nonzero resistivity at vanishing temperatures.
For $T\lesssim 0.1t\simeq \omega_g / 3$, electrons near the Fermi level thermally fluctuate so weakly that they cannot cross the gap to scatter off the second scattering mode~\cite{hartnoll22colloquium}.
Therefore, the scattering rate $-\Im\Sigma(0)$ at $T=0.1t$ saturates and is close to that at $T=0.05t$, leading to the lower resistivity plateau at $T \lesssim 0.1t$.
Similarly, the pseudogap renders the scattering rates $-\Im\Sigma(0)$ at $T\simeq 0.267t$ and $T\simeq 0.5t$ near and both close to that of the second scattering mode [$-\Im\Sigma(0.9t)$ at $T=0.05t$], forming the higher resistivity plateau at $0.26t\lesssim T \lesssim 0.5t$.
But because the pseudogap is not a genuine gap, the higher resistivity plateau is not as flat as the lower resistivity plateau (Fig.~\ref{fig:resis}, right panel).
For heavier doping, the Kondo resonance peak will get wider~\cite{pruschke95anomalous,georges96dynamical}, i.e., the (pseudo-) gap in $-\Im\Sigma(\omega)$ will become wider, resulting in wider resistivity plateaus.

\section{Conclusion}
In conclusion, we have calculated the Hall coefficient and longitudinal resistivity of the hole-doped bilayer Hubbard model and found that its transport properties are very different from those of the monolayer or single-orbital Hubbard model.
The Hall coefficient has a strong nonmonotonic dependence on temperature at low temperatures, and it can change sign four times for some range of doping.
The resistivity at low temperatures is not linear in temperature like that of strange metals or quadratic in temperature like that of Fermi liquids.
Rather, it exhibits two plateaus with a smooth crossover between them.
These anomalous transport behaviors can be traced back to the formation of coherent interlayer singlets, which scatter off gapped collective modes arising from short-range interlayer antiferromagnetic fluctuations.

Although we did not account for vertex corrections to the conductivity, the corresponding analysis for the single-band Hubbard model hints at the effect of the vertex corrections.
Inclusion of the vertex corrections shifts the longitudinal resistivity downward, but preserves its temperature dependence~\cite{vucicevic19conductivity,vranic20charge}.
Neither does it alter the trend of the Hall coefficient as a function of temperature~\cite{assaad95hall,lange99magnetotransport,wang20dc}.
Therefore, we do not expect the vertex corrections to give rise to resistivity behaviors qualitatively different from our conclusions.

\section{Acknowledgements}
This work was supported as part of the Computational Materials Sciences Program funded by the U.S. Department of Energy, Office of Science, Basic Energy Sciences, under Award No. DE-SC0020145.
JS was supported by the U.S. Department of Energy, Office of Basic Energy Sciences, under Grant No. DE-SC-0005042.
This research used resources of the National Energy Research Scientific Computing Center (NERSC), a U.S. Department of Energy Office of Science User Facility located at Lawrence Berkeley National Laboratory, operated under Contract No. DE-SC-0020145 using NERSC award BES-ERCAP0023632.

%\nocite{*}

\bibliography{bilayerhall_refs}% Produces the bibliography via BibTeX.

\end{document}